\documentclass[twocolumn,amsmath,amssymb,prl]{revtex4}

\usepackage{graphicx}
\usepackage{dcolumn}
\usepackage{bm}

\def\be{\begin{equation}}
\def\ee{\end{equation}}
\def\bea{\begin{eqnarray}}
\def\eea{\end{eqnarray}}

\def\OO#1{{\cal O}(c^{-#1})}

\begin{document}

\title{\textbf{TESTS OF LORENTZ INVARIANCE USING A MICROWAVE RESONATOR: AN UPDATE}}
\author{P. Wolf}
    \altaffiliation[On leave from: ]{Bureau International des Poids et Mesures, Pavillon de Breteuil, 92312 S\`evres Cedex, France}
\author{S. Bize}
\author{A. Clairon}
\author{G. Santarelli}
    \affiliation{BNM-SYRTE, Observatoire de Paris, 61 Avenue de l'Observatoire, 75014 Paris, France }
\author{A.N. Luiten}
\author{M.E. Tobar}
    \affiliation{University of Western Australia, School of Physics, Nedlands 6907 WA, Australia }


\maketitle

\textit{Abstract} - \textbf{The frequencies of a cryogenic sapphire oscillator and a hydrogen maser are compared to set new constraints on a possible violation of Lorentz invariance. We determine the variation of the oscillator frequency as a function of its orientation (Michelson-Morley test) and of its velocity (Kennedy-Thorndike test) with respect to a preferred frame candidate. We constrain the corresponding parameters of the Mansouri and Sexl test theory to $\delta - \beta + 1/2 \leq 3.4 \times 10^{-9}$ and $\beta - \alpha - 1 \leq 4.1 \times 10^{-7}$ which is of the same order as the best previous result for the former and represents a 50 fold improvement for the latter. These results correspond to an improvement of our previously published limits [Wolf P. et al., Phys. Rev. Lett. {\bf 90}, 6, 060402, (2003)] by about a factor 2. We describe the changes of the experiment, and show the new data that lead to that improvement.
}\\


\pagestyle{empty}

The Einstein equivalence principle (EEP) is at the heart of special and general relativity \cite{Will} and a cornerstone of modern physics. One of the constituent elements of EEP is Local Lorentz invariance (LLI) which, loosely stated, postulates that the outcome of any local test experiment is independent of the velocity of the freely falling apparatus (the fundamental hypothesis of special relativity). The central importance of this postulate in modern physics has motivated tremendous work to experimentally test LLI \cite{Will}. Additionally, nearly all unification theories (in particular string theory) violate the EEP at some level \cite{Damour1} which further motivates experimental searches for such violations of the universality of free fall \cite{Damour2} and of Lorentz invariance \cite{Kosto1,Kosto2}.

	The vast majority of modern experiments that test LLI rely essentially on the stability of atomic clocks and macroscopic resonators \cite{Brillet,KT,Hils,Schiller,Wolf}, therefore improvements in oscillator technology have gone hand in hand with improved tests of LLI. Our experiment is no exception, the large improvement being a direct result of the excellent stability of our cryogenic sapphire oscillator. Additionally its operation at a microwave frequency allows a direct comparison to a hydrogen maser which provides a highly stable and reliable reference frequency.

    Numerous test theories that allow the modeling and interpretation of experiments that test LLI have been developed. Kinematical frameworks \cite{Robertson,MaS} postulate a simple parametrisation of the Lorentz transformations with experiments setting limits on the deviation of those parameters from their special relativistic values. A more fundamental approach is offered by theories that parametrise the coupling between gravitational and non-gravitational fields (TH$\epsilon\mu$ \cite{LightLee,Will,Blanchet} or $\chi$g \cite{Ni} formalisms) which allow the comparison of experiments that test different aspects of the EEP. Finally, formalisms based on string theory \cite{Damour1,Damour2,Kosto1} have the advantage of being well motivated by theories of physics that are at present the only candidates for a unification of gravity and the other fundamental forces of nature.

    Owing to their simplicity the kinematical frameworks of \cite{Robertson,MaS} have been widely used to model and interpret many previous experiments testing LLI \cite{Brillet,Hils,Schiller,Riis,WP}. In order to compare our results to those experiments we will follow this route in the present work (an analysis of our experiment in the light of other test theories being relegated to a future publication). Those frameworks postulate generalized transformations between a preferred frame candidate $\Sigma(T,{\bf X})$ and a moving frame ${\rm S}(t,{\bf x})$ where it is assumed that in both frames coordinates are realized by identical standards (e.g. hydrogen masers for the time coordinates and sapphire rods for the length coordinates in our case). We start from the transformations of \cite{MaS} (in differential form) for the case where the velocity of S as measured in $\Sigma$ is along the positive X-axis, and assuming Einstein synchronization in S (we will be concerned with signal travel times around closed loops so the choice of synchronization convention can play no role):

\bea
dT &=& \frac{1}{a}\left(dt+\frac{vdx}{c^2}\right) \nonumber\\ 
dX &=& \frac{dx}{b}+\frac{v}{a}\left(dt+\frac{vdx}{c^2}\right)\\
dY &=& \frac{dy}{d}\nonumber \\
dZ &=& \frac{dz}{d}\nonumber\\ \nonumber
\eea
with $c$ the velocity of light in vacuum in $\Sigma$. Using the usual expansion of the three parametrs $(a \approx 1+\alpha{v^2/c^2} + \OO4; b \approx 1+\beta{v^2/c^2} + \OO4; d \approx 1+\delta{v^2/c^2} + \OO4)$, setting $c^2dT^2=dX^2+dY^2+dZ^2$ in $\Sigma$, and transforming according to (1) we find the coordinate travel time of a light signal in S:

\bea
dt &=& \frac{dl}{c}\left\lbrack 1-\left(\beta -\alpha -1 \right)\frac{v^2}{c^2} \right. \\
&-& \left. \left(\frac{1}{2}-\beta +\delta \right){\rm sin}^2\theta\frac{v^2}{c^2}\right\rbrack +\OO4 \nonumber
\eea
where $dl = \sqrt{dx^2+dy^2+dz^2}$ and $\theta$ is the angle between the direction of light propagation and the velocity {\bf v} of S in $\Sigma$. In special relativity $\alpha = -1/2; \beta = 1/2; \delta = 0$ and (1) reduces to the usual Lorentz transformations. Generally, the best candidate for $\Sigma$ is taken to be the frame of the cosmic microwave background (CMB) \cite{Fixsen,Lubin} with the velocity of the solar system in that frame taken as $v_\odot \approx 377$ km/s, decl. $\approx -6.4 ^\circ $, $RA \approx 11.2$h.

    Michelson-Morley type experiments \cite{MM,Brillet} determine the coefficient $P_{MM} = (1/2-\beta +\delta)$ of the direction dependent term. For many years the most stringent limit on that parameter was $|P_{MM}| \leq 5 \times 10^{-9}$ determined over 23 years ago in an outstanding experiment \cite{Brillet}. Our experiment confirms that result with roughly equivalent uncertainty $(3.4 \times 10^{-9})$. Recently an improvement to $|P_{MM}| \leq 1.3 \times 10^{-9}$ has been reported \cite{Muller}. Kennedy-Thorndike experiments \cite{KT,Hils,Schiller} measure the coefficient $P_{KT} = (\beta -\alpha -1)$ of the velocity dependent term. The most stringent limit \cite{Schiller} on $|P_{KT}|$ has been recently improved from \cite{Hils} by a factor 3 to $|P_{KT}| \leq 2.1 \times 10^{-5}$. We improve this result by a factor of 50 to $|P_{KT}| \leq 4.1 \times 10^{-7}$. Finally clock comparison and Doppler experiments \cite{Riis,Grieser,WP} measure $\alpha$, currently limiting it to $|\alpha + 1/2| \leq 8 \times 10^{-7}$. The three types of experiments taken together then completely characterize any deviation from Lorentz invariance in this particular test theory.

    Our cryogenic oscillator consists of a sapphire crystal of cylindrical shape operating in a whispering gallery mode (see fig. 1 for a schematic drawing and \cite{Chang,Mann} for a detailed description).

\begin{figure}[htb]
\includegraphics[height=6.5cm,width=8.5cm]{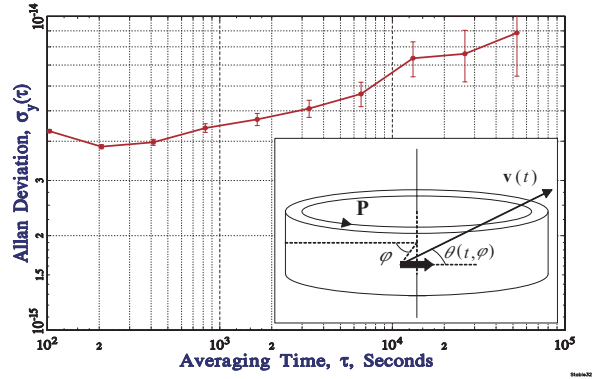}
\caption{Typical relative frequency stability of the CSO - H-maser difference after removal of a linear frequency drift. The inset is a schematic drawing of the cylindrical sapphire oscillator with the Poynting vector $\bf P$ in the whispering gallery (WG) mode, the velocity ${\bf v}(t)$ of the cylinder with respect to the CMB, and the relevant angles for a photon in the WG mode.
} \label{fig:clocks}
\end{figure}

Its coordinate frequency can be expressed by  $\nu = m/t_c$ where $t_c$ is the coordinate travel time of a light signal around the circumference of the cylinder (of radius $r$) and $m$ is a constant. From (2) the relative frequency difference between the sapphire oscillator and the hydrogen maser (which realizes coordinate time in S \cite{masercom}) is

\bea
\frac{\Delta \nu (t)}{\nu_0} &=& P_{KT}\frac{v(t)^2}{c^2} \\
&+& P_{MM}\frac{v(t)^2}{c^2}\frac{1}{2\pi}\int_0^{2 \pi}{\rm sin}^2\theta (t,\varphi ) d\varphi +\OO3 \nonumber
\eea
where $\nu_0 = m/(2\pi r/c)$, $v(t)$ is the (time dependent) speed of the lab in $\Sigma$, and $\varphi$ is the azimuthal angle of the light signal in the plane of the cylinder. The periodic time dependence of $v$ and $\theta$ due to the rotation and orbital motion of the Earth with respect to the CMB frame allow us to set limits on the two parameters in (3) by adjusting the periodic terms of appropriate frequency and phase (see \cite{Mike} for calculations of similar effects for several types of oscillator modes). Given the limited durations of our data sets ($\leq$ 15 days) the dominant periodic terms arise from the Earth's rotation, so retaining only those we have ${\bf v}(t) = {\bf u}+{\bf \Omega} \times {\bf R}$ with ${\bf u}$ the velocity of the solar system with respect to the CMB, ${\bf \Omega}$ the angular velocity of the Earth, and ${\bf R}$ the geocentric position of the lab. We then find after some calculation

\bea
\Delta \nu / \nu_0 = &P_{KT}(H{\rm sin}\lambda ) + P_{MM}(A{\rm cos}\lambda + B{\rm cos}(2\lambda) \nonumber \\
&+ C{\rm sin}\lambda+D{\rm sin}\lambda{\rm cos}\lambda+E{\rm sin}\lambda{\rm cos}(2\lambda)) \nonumber \\
\eea
where $\lambda =\Omega t + \phi$, and A-E and $\phi$ are constants depending on the latitude and longitude of the lab $(\approx 48.7 ^\circ$N and $2.33 ^\circ$E for Paris). Numerically $H \approx -2.6 \times 10^{-9}$, $A \approx -8.8 \times 10^{-8}$, $B \approx 1.8 \times 10^{-7}$, C-E of order $10^{-9}$. We note that in (4) the dominant time variations of the two combinations of parameters are in quadrature and at twice the frequency which indicates that they should decorelate well in the data analysis allowing a simultaneous determination of the two (as confirmed by the correlation coefficients below). Adjusting this simplified model to our data we obtain results that differ by less than 10\% from the results presented below that were obtained using the complete model ((3) including the orbital motion of the Earth).

	The cryogenic sapphire oscillator (CSO) is compared to a commercial (Datum Inc.) active hydrogen maser whose frequency is also regularly compared to caesium and rubidium atomic fountain clocks in the laboratory \cite{Bize}. The CSO resonant frequency at 11.932 GHz is compared to the 100 MHz output of the hydrogen maser. The maser signal is multiplied up to 12 GHz of which the CSO signal is subtracted. The  remaining $\approx$ 67 MHz signal is mixed to a synthesizer signal at the same frequency and the low frequency beat at $\approx$ 64 Hz is counted, giving access to the frequency difference between the maser and the CSO. The instability of the comparison chain has been measured and does not exceed a few parts in $10^{16}$. The typical stability of the measured CSO - maser frequency after removal of a linear frequency drift is shown in fig. 1. Since January 2003 we have implemented an active temperature control of the CSO room and changed some of the electronics. As a result the diurnal and semi-diurnal temperature variations during measurement runs ($\approx$ 2 weeks) were typically $\approx 0.1^\circ$ C in amplitude with best values $\leq 0.04^\circ$, and longer and more reliable data sets became available.

    Our experimental data consists of two blocks, before and after the Jan. 2003 upgrade. The first block consists of eight sets of measured values of $\Delta \nu / \nu_0$ taken between Nov. 2001 and Oct. 2002. Our previously published results \cite{Wolf} are based on the first seven of these (3 to 10 days individual lengths, 37 days in total). The second includes 5 data sets taken between Jan. 2003 and Apr. 2003 with significantly longer lengths (6 to 15 days, 56 days in total) and better temperature stability. The sampling times $\tau _0$ are generally 100 s except for the first two sets for which $\tau_0 = 12$ s and $5000$ s respectively. To analyze our data we simultaneously adjust an offset and a rate (natural frequency drift, typically $\approx 2 \times 10^{-18}$ s$^{-1}$) per data set and the two parameters of the model (3). In the model (3) we take into account the rotation of the Earth and the Earth's orbital motion, the latter contributing little as any constant or linear terms over the durations of the individual data sets are absorbed by the adjusted offsets and rates.

    When carrying out an ordinary least squares (OLS) adjustment we note that the residuals have a significantly non-white behavior as one would expect from the slope of the Allan deviation of fig. 1. The power spectral density (PSD) of the residuals when fitted with a power law model of the form $S_y(f)=kf^\mu$ yields typically $\mu \approx -1.5$ to $-2$. In the presence of non-white noise OLS is not the optimal regression method \cite{lss,Draper} as it can lead to significant underestimation of the parameter uncertainties \cite{lss}.

    An alternative method is weighted least squares (WLS) \cite{Draper} which allows one to account for non-random noise processes in the original data by pre-multiplying both sides of the design equation (our equation (3) plus the offsets and rates) by a weighting matrix containing off diagonal elements. To determine these off diagonal terms we first carry out OLS and adjust the $S_y(f)=kf^\mu$ power law model to the PSD of the post-fit residuals determining a value of $\mu$ for each data set. We then use these $\mu$ values to construct a weighting matrix following the method of fractional differencing described, for example, in \cite{lss}. Figure 2. shows the resulting values of the two parameters ($P_{KT}$ and $P_{MM}$) for each individual data set. As previously reported \cite{Wolf} a global WLS fit to the first 7 data sets yields $|P_{MM}| = (1.5\pm 3.1) \times 10^{-9}$ and $|P_{KT}| = (-3.1\pm 3.7) \times 10^{-7}$ ($1\sigma$ uncertainties), with the correlation coefficient between the two parameters less than 0.01 and all other correlation coefficients $< 0.07$. Because of the qualitative difference of the data we decided to treat the recent data (since Jan. 2003) independently, the global WLS adjustment for those 5 data sets yielding $|P_{MM}| = (5.5\pm 2.5) \times 10^{-9}$ and $|P_{KT}| = (1.2\pm 3.4) \times 10^{-7}$, again with insignificant correlation coefficients.

\begin{figure}[htb]
\includegraphics[height=8cm,width=9cm]{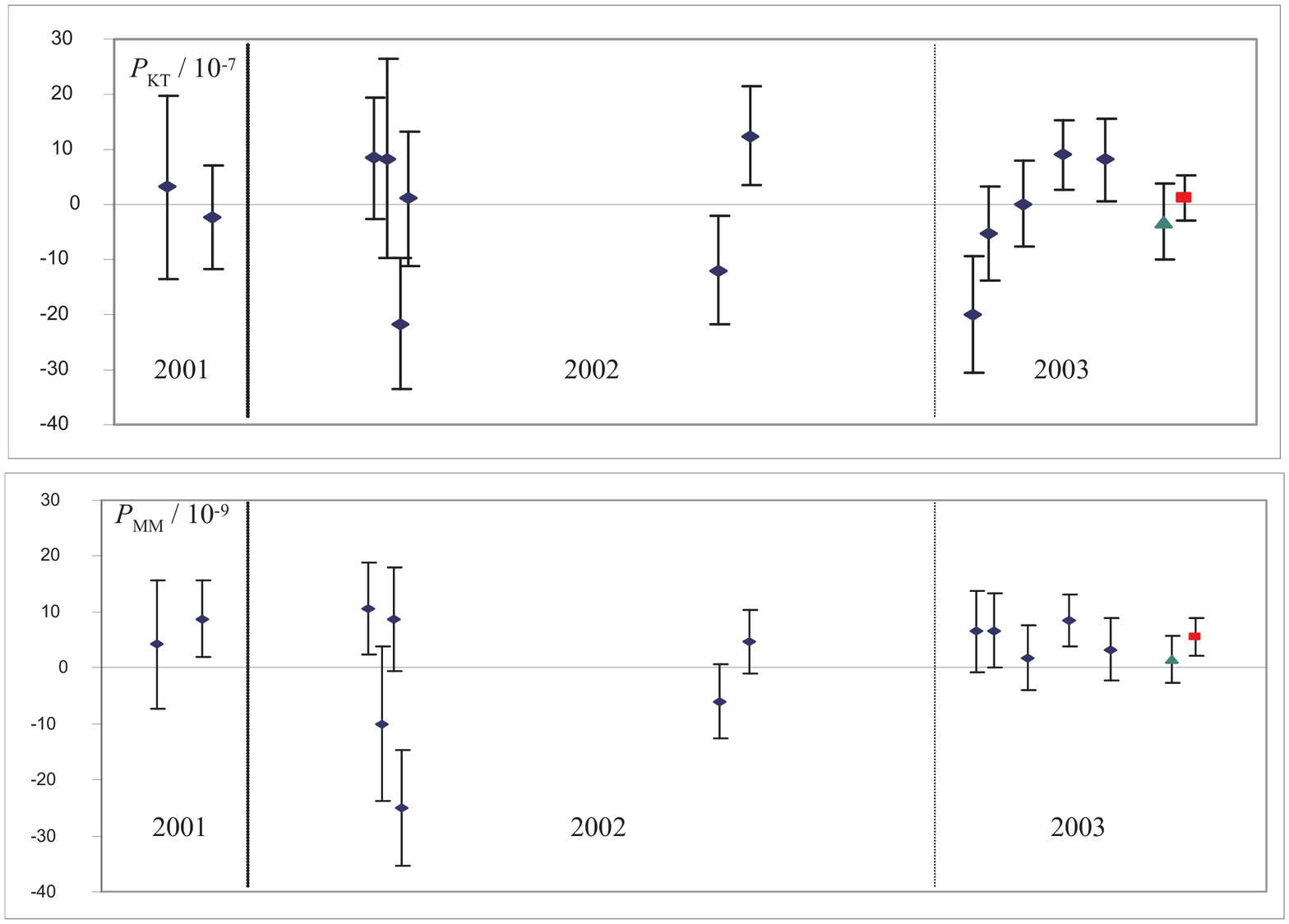}
\caption{Values of the two parameters ($P_{KT}$ and $P_{MM}$) from a fit to each individual data set (blue diamonds), to the first seven data sets (green triangle, as previously published \cite{Wolf}), and to the last five data sets (red square, data since upgrade). The error bars indicate the combined uncertainties from statistics and systematic effects (see text for details).} \label{fig:clocks}
\end{figure}

	Systematic effects at diurnal or semi-diurnal frequencies with the appropriate phase could mask a putative sidereal signal. The statistical uncertainties of $P_{MM}$ and $P_{KT}$ obtained from the WLS fit above correspond to sidereal and semi-sidereal terms (from (4)) of $\approx 1 \times 10^{-15}$ and $\approx 5 \times 10^{-16}$ respectively so any systematic effects exceeding these limits need to be taken into account in the final uncertainty. We expect the main contributions to such effects to arise from temperature, pressure and magnetic field variations that would affect the hydrogen maser, the CSO and the associated electronics, and from tilt variations of the CSO which are known to affect its frequency. Measurements of the tilt variations of the CSO show amplitudes of 4.6 $\mu$rad and 1.6 $\mu$rad at diurnal and semi-diurnal frequencies. To estimate the tilt sensitivity we have intentionally tilted the oscillator by $\approx$ 5 mrad off its average position which led to relative frequency variations of $\approx 3 \times 10^{-13}$ from which we deduce a tilt sensitivity of $\approx 6 \times 10^{-17} \mu$rad$^{-1}$. This value corresponds to a worst case scenario as we expect a quadratic rather than linear  frequency variation for small tilts around the vertical. Even with this pessimistic estimate diurnal and semi-diurnal frequency variations due to tilt do not exceed $3 \times 10^{-16}$ and $1 \times 10^{-16}$ respectively and are therefore negligible with respect to the statistical uncertainties. 

Before January 2003 temperature measurements of the CSO lab taken during some of the experimental runs show room temperature variations with amplitudes of 0.3 $^\circ$C and 0.1 $^\circ$C for the diurnal and semi-diurnal components which are reduced to 0.04 $^\circ$C and 0.01 $^\circ$C on the actively temperature controlled electronics panel. In december 2002 we implemented an active temperature stabilisation inside an isolated volume ($\approx 15 {\rm m}^3$) that included the CSO and all the associated electronics. The temperature was measured continously in two fixed locations (behind the electronics rack and on top of the dewar) the latter being the reference point for the stabilisation servo loop. Using this system the diurnal/semi-diurnal temperature variations observed were 0.1/0.08 $^\circ$C in amplitude, for the worst data set and the measurement behind the electronics rack (on the dewar the fluctuations were always significantly smaller). The quoted numbers are the worst performance observed over all available data sets since Jan. 2003, in best cases they remained as low as 0.04/0.04 $^\circ$C. The hydrogen maser is kept in a dedicated, environmentally controlled clock room. Measurements of magnetic field, temperature and atmospheric pressure in that room and the maser sensitivities as specified by the manufacturer allow us to exclude any systematic effects on the maser frequency that would exceed the statistical uncertainties above. When heating and cooling the CSO lab by $\approx 3^\circ$C we see frequency variations of $\approx 5 \times 10^{-15}$ per $^\circ$C. This is also confirmed when we induce a large sinusoidal temperature variation ($\approx 1.5 ^\circ$C amplitude). From the temperature measurements during the experimental runs before Jan. 2003 we therefore deduce a total diurnal/semi-diurnal effect of $\approx 1.5 \times 10^{-15}$ and $\approx 0.5 \times 10^{-15}$ respectively. For the more recent data (taking the worst temperature variations observed) we obtain $\approx 0.5/0.4 \times 10^{-15}$. We assume the pessimistic scenario where all of the diurnal and semi-diurnal effect is present at the neighboring sidereal and semi-sidereal frequencies which leads (from (3)) to uncertainties from systematic effects of $\pm 5.8 \times 10^{-7}$ on $P_{KT}$ and $\pm 2.8 \times 10^{-9}$ on $P_{MM}$ for the old data and $\pm 2.3 \times 10^{-7}$ respectively $\pm 2.2 \times 10^{-9}$ for the data since Jan. 2003. We note that the phase of the perturbing systematic signal will vary over the course of our measurements due to natural causes (meteorology, daytime changes etc.) and to the sidereal/diurnal frequency difference so our final uncertainties are the quadratic sums of the above values and the statistical uncertainties from the WLS adjustment: $|P_{MM}| \leq 3.4 \times 10^{-9}$ and $|P_{KT}| \leq 4.1 \times 10^{-7}$ for the new data.

In summary, we have reported an experimental test of Lorentz invariance that simultaneously constrains two combinations of the three parameters of the Mansouri and Sexl test theory (previously measured individually by Michelson-Morley and Kennedy-Thorndike experiments). Using our recent results (after Jan. 2003) we obtain $|\delta - \beta + 1/2| = 5.5(2.5)(2.2) \times 10^{-9}$ which is of the same order as the best previous results \cite{Brillet,Muller}, and $|\beta - \alpha - 1| = 1.2(3.4)(2.3)\times 10^{-7}$ which improves the best present limit \cite{Schiller} by a factor of 50 (the first bracket indicates the uncertainty from statistics the second from systematic effects). The measured deviation of $|\delta - \beta + 1/2|$ from zero is significant at the $1\sigma$ level (67 \% confidence) and could indicate a deviation from Lorentz invariance. However, at the time being, we consider this to be most likeley a statistical artefact or an overlooked or insufficiently evaluated systematic effect. The experiment is running continuously and we expect that future data will help to clarify this situation. As a result of our experiment the Lorentz transformations are confirmed in this particular test theory with an overall uncertainty of $\leq 8 \times 10^{-7}$ limited now by the determination of $\alpha$ from Doppler and clock comparison experiments \cite{Riis,WP}. This is likely to be improved in the coming years by experiments such as ACES (Atomic Clock Ensemble in Space \cite{ACES}) that will compare ground clocks to clocks on the international space station aiming at a 10 fold improvement on the determination of $\alpha$.

\begin{center}
ACKNOWLEDGEMENTS
\end{center}
Helpful discussions with Christophe Salomon and G\'erard Petit are gratefully acknowledged as well as financial support by the Australian Research Council and CNES. P.W. is supported by CNES research grant 793/02/CNES/XXXX.


\begin{thebibliography}{99}
\bibitem{Will} Will C.M., {\it Theory and Experiment in Gravitational Physics, revised edition},
Cambridge U. Press, (1993).
\bibitem{Damour1} Damour T., gr-qc/9711060 (1997).
\bibitem{Damour2} Damour T. and Polyakov A.M., Nucl.Phys. {\bf B423}, 532, (1994).
\bibitem{Kosto1} Colloday D. and Kostelecky V.A., Phys. Rev. {\bf D55}, 6760, (1997).
\bibitem{Kosto2} Bluhm R. et al., Phys. Rev. Lett. {\bf 88}, 9, 090801, (2002).
\bibitem{Brillet} Brillet A. and Hall J.L., Phys. Rev. Lett. {\bf 42}, 9, 549, (1979).
\bibitem{KT} Kennedy R.J. and Thorndike E.M., Phys. Rev. {\bf B42}, 400, (1932).
\bibitem{Hils} Hils D. and Hall J.L., Phys. Rev. Lett., {\bf 64}, 15, 1697, (1990).
\bibitem{Schiller} Braxmaier C. et al., Phys. Rev. Lett. {\bf 88}, 1, 010401, (2002).
\bibitem{Wolf} Wolf P. et al., Phys. Rev. Lett. {\bf 90}, 6, 060402, (2003).
\bibitem{Robertson} Robertson H.P., Rev. Mod. Phys. {\bf 21}, 378 (1949).
\bibitem{MaS} Mansouri R. and Sexl R.U., Gen. Rel. Grav. {\bf 8}, 497, 515, 809, (1977).
\bibitem{LightLee} Lightman A.P. and Lee D.L., Phys. Rev. {\bf D8}, 2, 364, (1973).
\bibitem{Blanchet} Blanchet L., Phys. Rev. Lett. {\bf 69}, 4, 559, (1992).
\bibitem{Ni} Ni W.-T., Phys. Rev. Lett. {\bf 38}, 301, (1977).
\bibitem{Riis} Riis E. et al., Phys. Rev. Lett. {\bf 60}, 81, (1988).
\bibitem{WP} Wolf P. and Petit G., Phys. Rev. {\bf A56}, 6, 4405, (1997).
\bibitem{Fixsen} Fixsen D.J. et al., Phys. Rev. Lett. {\bf 50}, 620, (1983).
\bibitem{Lubin} Lubin et al., Phys. Rev. Lett. {\bf 50}, 616, (1983).
\bibitem{MM} Michelson A.A. and Morley E.W., Am. J. Sci., {\bf 34}, 333, (1887).
\bibitem{Muller} Schiller S. et al., {\it Proc. 6th Symp. on Frequency Standards and Metreology}, Gill P. ed, World Scientific, (2002).
\bibitem{Grieser} Grieser R. et al., Appl. Phys. {\bf B59}, 127, (1994).
\bibitem{Chang} Chang S., Mann A.G. and Luiten A.N., Electron. Lett. {\bf 36}, 5, 480, (2000).
\bibitem{Mann} Mann A.G., Chang S. and Luiten A.N., IEEE Trans. Instrum. Meas. {\bf 50}, 2, 519, (2001).
\bibitem{masercom} We assume throughout the paper that LPI is sufficiently verified so that the variation of the maser frequency due to the diurnal variation of the local gravitational potential is negligible. Indeed the results of \cite{Bauch} imply that such variations should not exceed 2 parts in $10^{-17}$ which is significantly below our noise level.
\bibitem{Bauch} Bauch A., Weyers S., Phys. Rev. {\bf D65}, 081101, (2002) 
\bibitem{Mike} Tobar M.E. et al., Phys. Lett. {\bf A300}, 33, (2002).
\bibitem{Bize} Bize S. et al., Proc. 6th Symp. on Freq. Standards and Metrology, World Scientific, (2002).
\bibitem{lss} Schmidt L.S., Metrologia {\bf 40}, in press, (2003).
\bibitem{Draper} Draper N.R. and Smith H., {\it Applied Regression Analysis}, Wiley, (1966).
\bibitem{ACES} Salomon C., et al., C.R. Acad. Sci. Paris, {\bf 2}, 4, 1313, (2001).

\end{thebibliography}
\end{document}